\documentclass{hitec}
\settextfraction{0.95}

\title{Photodetectors and front-end electronics for the LHCb RICH upgrade}

\author{L. Cassina, on behalf of the LHCb RICH collaboration}

\company{University and INFN of Milano-Bicocca}
\company{email: lorenzo.cassina@mib.infn.it}

\usepackage{hyperref} 
\usepackage{array}
\usepackage{tabularx}
\usepackage{listings}
\usepackage{graphicx}
\usepackage{caption}
\usepackage{float}
\usepackage{multirow}

\usepackage{amsmath}
\usepackage{amssymb,latexsym}
\usepackage{sansmath}\sansmath

\usepackage{bytefield}

\begin{document}
\maketitle

\begin{abstract}
The RICH detectors of the LHCb experiment provide identification of hadrons produced in high energy proton-proton collisions in the LHC at CERN over a wide momentum range (2 to 100 GeV/c). Cherenkov light is collected on photon detector planes sensitive to single photons. 
The RICH will be upgraded (in 2019) to read out every bunch crossing, at a rate of 40 MHz. The current hybrid photon detectors (HPD) will be replaced with multi-anode photomultiplier tubes (customisations of the Hamamatsu R11265 and the H12699 MaPMTs). These 8$\times$8 pixel devices meet the experimental requirements thanks to their small pixel size, high gain, negligible dark count rate ($\sim$50 Hz/cm$^2$) and moderate cross-talk. The measured performance of several tubes is reported, together with their long-term stability. A new 8-channel front-end chip, named CLARO, has been designed in 0.35 $\mu$m CMOS AMS technology for the MaPMT readout. The CLARO chip operates in binary mode and combines low power consumption (\hbox{$\sim$1 mW/Ch}), wide bandwidth (baseline restored in $\leq$25 ns) and radiation hardness. A 12-bit digital register permits the optimisation of the dynamic range and the threshold level for each channel and provides tools for the on-site calibration. The design choices and the characterization of the electronics are presented.
\end{abstract}




\section{The Upgraded RICH for the LHCb experiment}\label{sec:Introduction}

LHCb \cite{LHCb} is one of the four large detectors operating at the LHC at CERN and it is mainly devoted to CP violation measurements and to the search for new physics in rare decays of beauty and charm hadrons produced in proton-proton collisions. One of the key detector features is the capability to identify particles ($\pi$, $K$ and $p$) over a wide momentum range (2-100 GeV/c) in order to distinguish final decay states of similar topology, reduce the combinatorial background and efficiently tag the particle flavour. These goals are achieved by exploiting two ring-imaging Cherenkov \cite{RICH} (RICH) stations, named \hbox{RICH 1} and \hbox{RICH 2}, located, respectively, upstream and downstream the LHCb dipole magnet.

An upgrade of the LHCb detector will take place in 2019-2020 to run at higher luminosity and operate at 40 MHz read-out rate \cite{LHCbUpgrade}. In particular, the RICH detector will be updated and the currently used hybrid photon detectors (HPDs) will be replaced by multi-anode photomultiplier tubes (MaPMTs) coupled with external wide-bandwidth read-out electronics \cite{TDR2014}. 
The photosensitive planes of the RICH detector are designed with a modular structure where the smallest units are called elementary cell (EC). Two EC models (EC-R houses four 1$\times$1 inch$^2$ MaPMTs, EC-H houses a single 2$\times$2 inch$^2$ MaPMT) were designed so that the photosensor pixel size can be chosen such that neither occupancy nor spatial resolution are limiting the performance. The system will be composed of $\sim$700 EC-R covering the RICH 1 and the central part of the RICH 2 photosensitive planes, while $\sim$400 EC-H will be used in the peripheral areas of the RICH 2 detector. This design ensures significant reduction of the costs with a negligible degradation of the overall PID performance. The final system will consist of $\sim$3100 MaPMTs and $\sim$2$\cdot$10$^5$ channels. In order to manage such a large number of channels and achieve the best performance from each pixel, the EC will be equipped with procedures for the on-site calibration of both the MaPMT and the electronic read-out chain.

An overview of the EC will be provided in section \ref{sec:EC} and the scheme for the rejection of spurious signals will be described in section \ref{sec:Noise}. Finally, the procedures developed for the individual channel calibration are described in section \ref{sec:Calibration}.

\section{Elementary cell}\label{sec:EC}

\begin{figure}[!ht]
\centering
\includegraphics[height=0.35\textwidth]{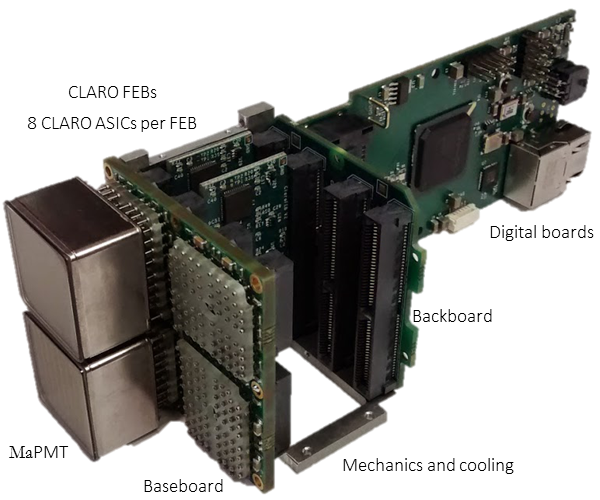}
\caption{Half-instrumented EC-R}
\label{fig:EC}
\end{figure}

An isometric view of a half-mounted EC-R for the upgraded LHCb RICH detector is shown in Fig.\ref{fig:EC}. The incoming Cherenkov photons produced by a particle moving faster than light in the dielectric gas radiator hit the MaPMT surface. The Hamamatsu R13742 MaPMT (a customization of the R11265 tube \cite{R11265}) is used for the R-type cell, whilst the EC-H is equipped with the Hamamatsu R13743 device, a customization of the R12699 MaPMT \cite{H12700}. The photodetectors are mounted in a custom support, called baseboard (BsB), which hosts the voltage divider biasing the sensors and provides a cross-shaped thermal mass, which behaves as a low thermal impedance path driving the heat dissipated by the divider towards the metallic structure surrounding the EC.  The photocathode of the MaPMT converts the incident photons into electrons via the photoelectric effect, and the electrical signal is then amplified by a 10-stage dynode chain, so that $\sim$10$^6$ electrons per photon are collected at each anode. The electric signal is transmitted by the BsB to the front-end board (FEB), each hosting 8 CLARO ASICs. The CLARO \cite{CLARO} is a 8-channel chip designed in 0.35 $\mu$m CMOS technology from AMS\footnote{Austria Micro Systems, website \url{http://ams.com}}. It consists of a charge-sensitive preamplifier coupled to a fast discriminator output stage. The CLARO is able to fully recover the baseline level after each pulse in $\leq$25 ns, so that a 40 MHz read-out rate can be achieved, and ensures a low power consumption, of the order of 1 mW/channel, avoiding the need of a dedicated front-end cooling system. Each channel can be configured to apply one of four gain values and 64 threshold values (implemented as a 12-bit register). The digital block also enables and manages the calibration procedures described in section \ref{sec:Calibration}. In order to guarantee a reliable operation in the LHCb environment, the digital block was fabricated using a radiation-hard design \cite{Siviglia1}\cite{Siviglia2} and protected from single-event upset by adopting a triple modular redundancy architecture\cite{RadHard}. At the CLARO output a digital pulse is provided if the integrated charge collected at the MaPMT anode is larger than the desired trigger threshold. This pulse, interpreted as the detection of a photon hitting the pixel under study, passes through the backboard and reaches a digital board\footnote{The final design of the DB is currently in development. The DB shown in Fig.\ref{fig:EC} is a preliminary version.} (DB). The core of the DB is a FPGA capable of collecting the CLARO signals, sending the data out of the detector, configuring the operational parameters of the chip and interfacing to the remote control. Finally, the metallic structure surrounding the cell is coupled to the cooling system allowing the system to operate close to room temperature.
 
\section{Elementary Cell design}\label{sec:Noise}
Each digital pulse provided at the CLARO output and recorded by the DB is interpreted as the detection of a photoelectron. This might be from a photon emitted by a particle crossing the detector faster than the speed of light in the detector radiator or from a noise source.
The effects of the different sources of noise would be interpreted as hits only if they cause signals larger than the CLARO trigger threshold. 
The MaPMT performance and the design of the EC were thus specifically studied to  maximize both the signal detection efficiency and the rejection of spurious counts.

\begin{figure}[!ht]
\centering
\includegraphics[height=0.35\textwidth]{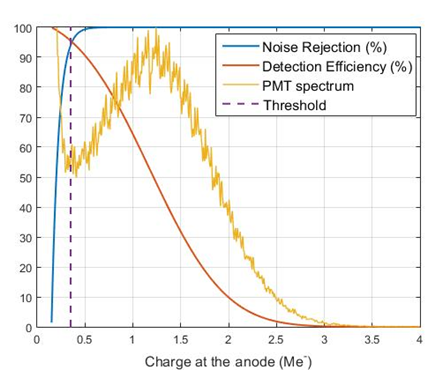}
\caption{Typical single photon spectrum (yellow), light signal detection efficiency (orange) and dark counts rejection efficiency (blue) as a function of the signal amplitude. The dashed line shows the trigger threshold}
\label{fig:Efficiency}
\end{figure}

Figure \ref{fig:Efficiency} shows a typical single photon spectrum of a MaPMT pixel. Assuming that the pedestal is only populated by spurious events while the single photon peak contains the signal of interest, the noise rejection and the signal detection efficiency as a function of the charge collected at the anode are superimposed in blue and orange respectively. By setting the trigger threshold in the valley between the two peaks, efficiencies of the order of 95\% can be obtained for both the signal detection and the noise rejection. Summarizing, a high signal over noise ratio can be obtained by setting the trigger threshold level pixel-by-pixel (as explained in section \ref{sec:Calibration}) and minimizing the rate of spurious counts above this level.

The main sources of spurious events in the LHCb RICH application are the spontaneous electron emission from either photocathode or dynodes via thermionic effect, the charge sharing effect between neighbouring pixels and the cross-talk due to the capacitive coupling between adjacent anodes. Dark counts due to thermionic emission from dynodes usually result in low amplitude signals, mainly located in the pedestal peak, and thus rejected. On the other hand, any electron emitted from the photocathode is amplified by the dynodes chain in the same way as the photon signal, so that these two contributions cannot be discriminated. At room temperature the dark count rate of the MaPMTs is $\leq$100 Hz/cm$^{2}$, totally negligible if compared to the signal rate ($\sim$10 MHz/pixel). The thermionic dark rate increases exponentially with the operating temperature, according to Richardson's law. So cooling is required. This requirement guided the design of BsB (equipped with a thermal mass) and CLARO (extremely low power consumption). The MaPMT should operate at 20-25 $^\circ$C in the final setup so that the thermionic emission rate is negligible with respect to the signal rate.

Charge sharing and cross-talk would induce spurious signals correlated with the Cherenkov photon signal rate. To ensure the rate of counts due to these effects is negligible, the amplitude of the induced signal must be lower than the trigger threshold in the associated pixel. To study the cross-talk within the MaPMT a dedicated setup was prepared, allowing illumination of a single pixel while acquiring simultaneously the waveform of the signals in all neighbouring pixels. 
\begin{figure}[!ht]
\centering
\includegraphics[height=0.35\textwidth]{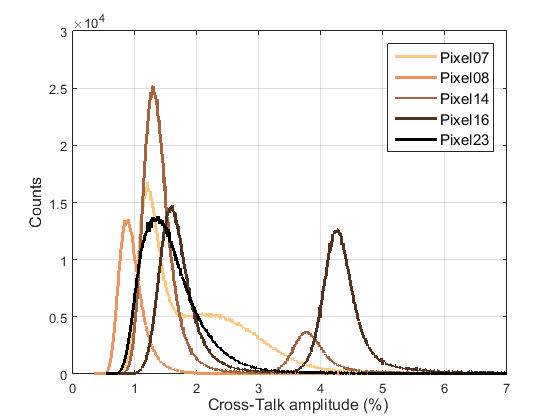}
\caption{Cross-talk amplitude distribution acquired while illuminating pixel 15 (H12699 MaPMT, HV=1 kV).}
\label{fig:CrossTalk}
\end{figure}

The ratio between the amplitudes of the signals recorded in the dark and illuminated pixels can be measured as a function of the inducing signal amplitude. Figure \ref{fig:CrossTalk} shows the cross-talk amplitude distribution acquired on a H12699 MaPMT, illuminating pixel 15 and recording the cross-talk signal in all the neighbouring pixels. The distributions in all the channels exhibit a peak at $\sim$1-2\% due to the capacitive coupling between the pixels, as proven by the bipolar shape of the signals acquired in the associated pixels \cite{H12700}. In adjacent pixels (referring to Fig.\ref{fig:CrossTalk}, pixels 14 and 16 are adjacent to the illuminated pixel 15) pulses with the same polarity as the inducing signal were observed. This signals are caused by the sharing of the charge among adjacent pixels during the multiplication chain and result in a peak in the cross-talk amplitude distribution at $\sim$5\%. Given these results, in the worst case condition, assuming the trigger threshold in each pixel of an EC has been set in the valley between the pedestal and the single photon peak, only 1\% of the single photon signals is expected to be large enough to induce a triggered spurious event in the associated pixel.

The effects described above are ascribed to the photosensors. However, the amplitude of the cross-talk signal is proportional to the stray capacitance between neighbouring electronic channels. Hence, the length of the electrical path connecting the MaPMT anodes to the CLARO input should be as short as possible. This consideration guided the choice to keep the number of channels per chip low, so that shorter connections, with a stray inter-channel capacitance of $\sim$0.2 pF, can be used. The BsB was also designed to fulfil this requirement and contributes $\sim$0.2 pF to the total capacitance between neighbouring channels. The contribution of the CLARO is negligible since a ground pin is placed at both sides of each input pin reducing the capacitive coupling between inputs. Summarizing, the total input capacitance between the inputs is of the order of $\sim$0.5 pF. 

In June 2016, beam tests were performed in the North Area of the Prevessin site at CERN to study the capability of four ECs to detect Cherenkov photons produced by 180 GeV/c pions and protons obtained from the SPS facility. The test conditions simulated the environment expected in the final setup. The reconstruction of the Cherenkov ring with a high resolution was achieved \cite{RICH2016-Carniti}. The test allows us to evaluate the spurious counts rate obtained when each of the $\sim$250 tested pixels have their trigger threshold set in the valley between the pedestal and the single photon peak. It turned out that the average spurious count rate amounts to $\sim$0.8\% of the signal rate. This result is compatible to the expectations assuming the MaPMT is the principal source of spurious events and neglecting the contribution due to the read-out system, thus validating the EC design choices.

\section{Calibration procedures}\label{sec:Calibration}

\begin{figure}[!ht]
\centering
\includegraphics[height=0.35\textwidth]{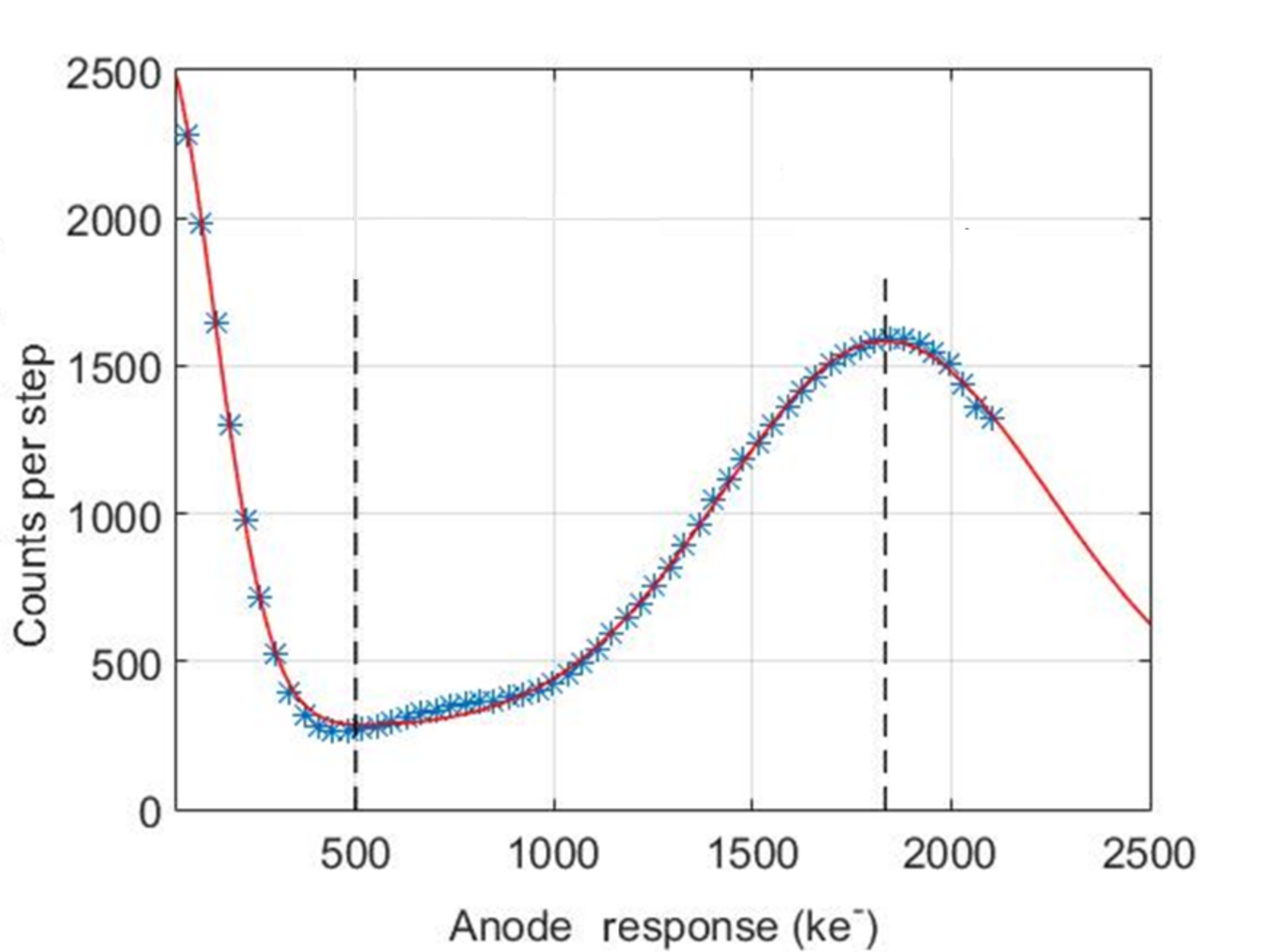}
\caption{Single photon peak acquired with a threshold scan. Pixel gain and optimized trigger threshold position are shown with dashed lines.}
\label{fig:Threshold}
\end{figure}

The previous sections have demonstrated the importance of setting a specific trigger threshold in each channel of the system in order to provide high signal detection efficiency and spurious counts rejection. From Fig.\ref{fig:Efficiency} one can see that the noise rejection efficiency curve is quite steep around the threshold value, so that errors of the order of 100 ke$^-$ can degrade the purity of the triggered events down to $\sim$85\%. The CLARO was designed to provide a resolution in the threshold setting of 30 ke$^-$, more than adequate for this application. However, the gain spread among pixels of the same photosensor can amount to a factor 3, while the mean gain spread between various devices can only be reduced by adjusting the biasing high voltage. Moreover, the ageing of the MaPMT during the data taking can cause a gain loss up to $\sim$20\% in the most illuminated detector regions \cite{R11265}. Hence, the EC must be equipped with a procedure to measure, pixel-by-pixel, single photon spectra in order to compensate gain spread and drift. 
The method developed for the MaPMT calibration, called \emph{threshold scan}, relies on a stable illumination level, easily achievable both on test benches and with the experimental setup. The incident photon rate is measured at various trigger thresholds in the range of interest. By taking the derivative of this distribution with respect of the threshold value, the single-photon spectra can be obtained. Figure \ref{fig:Threshold} shows a single-photon spectrum acquired by performing a threshold scan over the available threshold range. Pixel gain and trigger threshold can be deduced from the single photon peak and valley position.

\begin{figure}[!ht]
\centering
\includegraphics[height=0.35\textwidth]{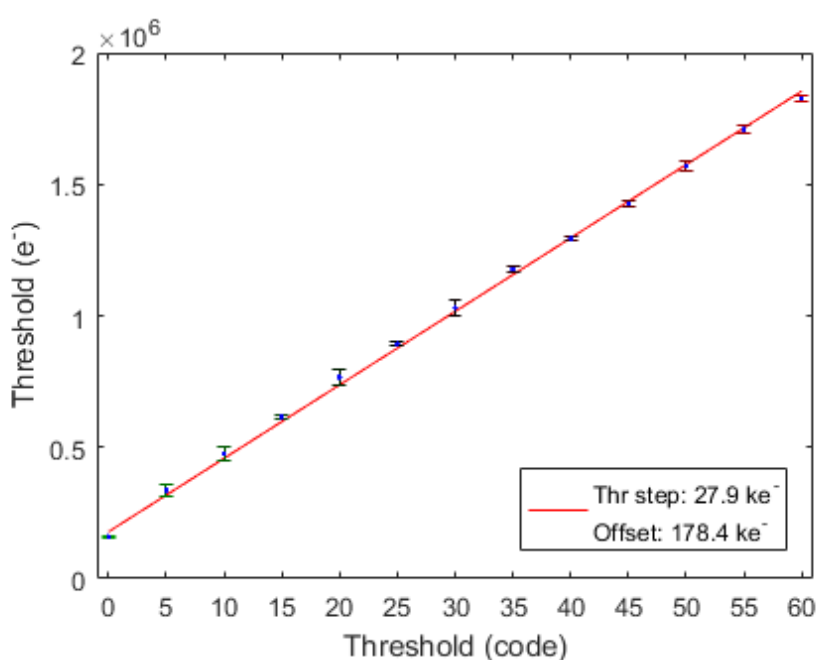}
\caption{Linear trend acquired by performing a DAC scan. This curve can be used to convert the trigger threshold code in equivalent electrons at the MaPMT anode.}
\label{fig:DAC}
\end{figure}

Variations in the production process can cause a spread between the threshold step of different channels ($\sigma$$\simeq$20\%, or \hbox{4.3 ke$^-$} at the maximum resolution) and an offset of $\sigma$=1.5 threshold step. In order to compensate these effects and precisely convert the threshold code to the equivalent input electrons, a calibration process called a \emph{DAC scan} was developed. An 8-bit Digital-to-Analog converter, currently located on the backboard and controlled via the FPGA, charges a 640 fF capacitor embedded in the CLARO.  As the capacitor discharges, a known charge is injected at the CLARO inputs. Fixing a certain threshold value, the S-curve relating the calibrating signal detection efficiency as a function of the DAC level amplitude can be measured. The position of the S-curve edge represents the trigger threshold amplitude, while the slope of the transition gives the noise \cite{TWEPP2015-Cassina}. By repeating this process for all the desired thresholds, the calibration shown in Fig.\ref{fig:DAC} is obtained. From this curve, threshold step and offset can be measured pixel-by-pixel, respectively from its slope and intercept, and the CLARO response linearity can be studied.

Both threshold and DAC scans were employed behaved during the beam test of June 2016\cite{RICH2016-Carniti} and satisfactory results were achieved.
In the final RICH setup, the process can be run in parallel for all the pixels. 

\section{Conclusion}\label{sec:Conclusion}

The LHCb RICH detector is planned to be upgraded in \hbox{2019-2020}. The upgrade consists of the replacement of the currently used HPDs with MaPMTs (R13742 and R13743 by Hamamatsu) read out by an external custom ASIC, called CLARO. The photosensitive planes of both RICH 1 and RICH 2 detectors will have a modular architecture where the elementary cell is the basic unit. The design choices aimed to ensure a high signal detection and noise rejection efficiency have been presented. Two calibration processes were developed for the compensation of the gain spread among pixels and for the calibration of the electronic chain in the final experimental setup. The operating principle of such procedures and the obtained results have been demonstrated.

\end{document}